\documentclass[a4paper,fleqn,usenatbib]{mnras}

\usepackage{newtxtext,newtxmath}

\usepackage[T1]{fontenc}
\usepackage{ae,aecompl}

\usepackage{graphicx}	
\usepackage{amsmath}	
\usepackage{amssymb}	
\usepackage{tikz}

\title[Observing external disc photoevaporation]{The Observational Anatomy of Externally Photoevaporating Planet-Forming Discs I: Atomic Carbon }

\author[Haworth \& Owen]
{\parbox{\textwidth}{Thomas J. Haworth$^{1}$\thanks{E-mail: \texttt{t.haworth@qmul.ac.uk}} and James E. Owen$^2$ 
}\vspace{0.4cm}\\
\parbox{\textwidth}{$^{1}$ Astronomy Unit, School of Physics and Astronomy, Queen Mary University of London, London E1 4NS, UK \\
$^{2}$ Astrophysics Group, Imperial College London, Blackett Laboratory, Prince
Consort Road, London SW7 2AZ, UK \\
}}

\begin{document}

\date{Accepted ???. Received ???; in original form ???}

\pagerange{\pageref{firstpage}--\pageref{lastpage}} \pubyear{2020}

\maketitle
\label{firstpage}

\begin{abstract}
We demonstrate the utility of CI as a tracer of photoevaporative winds that are being driven from discs by their ambient UV environment. Commonly observed CO lines only trace these winds in relatively weak UV environments and are otherwise dissociated in the wind at the intermediate to high UV fields that most young stars experience. However, CI traces unsubtle kinematic signatures of a wind in intermediate UV environments ($\sim1000$\,G$_0$) and can be used to place constraints on the kinematics and temperature of the wind.  In CI position-velocity diagrams external photoevaporation results in velocities that are faster than those from Keplerian rotation alone, as well as emission from quadrants of position-velocity space in which there would be no Keplerian emission. This is independent of viewing angle because the wind has components that are perpendicular to the azimuthal rotation of the disc. At intermediate viewing angles ($\sim 30-60^\circ$) moment 1 maps also exhibit a twisted morphology over large scales (unlike other processes that result in twists, which are typically towards the inner disc). CI is readily observable with ALMA, which means that it is now possible to identify and characterise the effect of external photoevaporation on planet-forming discs in intermediate UV environments. 
\end{abstract}


\section{introduction}
With the advent of ALMA, our view of planet-forming discs has been revolutionised in recent years. The picture of discs we had from early observations as smooth ``continuous'' structures \citep[e.g. as they appear in optical observations of discs in Orion,][]{1996AJ....111.1977M} is changed. We now know from (sub)millimetre continuum images that, at least in dust grains of order a millimetre in size, substructures such as rings and spirals are common \citep{2015ApJ...808L...3A, 2018ApJ...869L..41A} and can be induced early in the disc lifetime at $<1\,$Myr \citep{2015ApJ...808L...3A}. These observations have rightly fed back into and now dominate substantial parts of our theoretical efforts as a community to understand planet formation \citep[e.g.][]{2015MNRAS.453L..73D, 2018ApJ...869L..45B, 2018ApJ...869L..46D, 2018ApJ...869L..47Z}. However, there is a caveat. These recent resolved observations have only been of discs at distances within $\sim170$\,pc of the Sun, and as such are all in low mass stellar clusters where the lack of massive stars means that the UV environment is weak.  Most star/discs will be in higher UV radiation environments \citep{2008ApJ...675.1361F, 2020MNRAS.491..903W} for which we have statistical observational evidence that the disc mass \citep{2014ApJ...784...82M, 2017AJ....153..240A}, size \citep{2018ApJ...860...77E} and lifetime \citep{2016arXiv160501773G} can all be much smaller. A smaller disc radius can also redistribute angular momentum on shorter timescales, leading to an enhanced accretion rate \citep{2017MNRAS.468.1631R}.  Although the nearby systems where we are resolving substructure are certainly excellent laboratories for exploring and understanding disc processes, we may be trying to understand planet formation and the link to planet populations using a subset of discs in uncommon environments.

Theoretically there is currently a consensus that the environmental radiation field will influence the disc evolution \citep[e.g.][]{2000ApJ...539..258R,2004ApJ...611..360A, 2011PASP..123...14H, 2013ApJ...774....9A, 2016MNRAS.457.3593F, 2018MNRAS.478.2700W, 2020MNRAS.491..903W, 2017MNRAS.468L.108H, 2018MNRAS.475.5460H, 2018MNRAS.481..452H, 2019MNRAS.485.3895H, 2019MNRAS.490.5678C}, which is supported by the aforementioned observed statistical trends in disc properties as a function of UV field strength. These modelling efforts have shown that external photoevaporation usually dominates over the environmental impact of gravitational encounters \citep{2001MNRAS.325..449S, 2018MNRAS.478.2700W} and influences the disc evolution (and potentially planet formation) by
\begin{enumerate}
\item Limiting the mass reservoir in the disc, which constrains the required planet formation efficiency \citep[e.g.][]{2018MNRAS.475.5460H, 2019MNRAS.490.5678C, 2020MNRAS.491..903W}. Note that this is important for planet formation at any radius in the disc, since the mass of close-in planets requires the migration of material from elsewhere in the disc. We cannot therefore ignore external photoevaporation on the premise that it affects the outer disc and most planets are being detected at smaller radii. Ultimately this also constrains the disc lifetime and hence timescale for planet formation \citep[e.g.][]{2019MNRAS.490.5678C, 2020MNRAS.491..903W}
\item Truncating the disc, which not only limits the locations for possible planet formation, but also lowers the viscous timescale, hence increasing the mass accretion rate and lowering the disc lifetime \citep[e.g.][]{2007MNRAS.376.1350C, 2017MNRAS.468L.108H, 2017MNRAS.468.1631R}. 
\item Heating the disc, which can suppress cold Jupiter populations \citep{2018MNRAS.474..886N}
\end{enumerate}
So the expectation from theory summarised above is qualitatively consistent with the observed statistical evidence for UV environment affecting disc properties. 

This combined observational evidence and theoretical expectation of the key role played by radiation environment exemplifies its importance, however external photoevaporation has proven difficult to understand and fully gauge the impact of in reality. One key obstacle to understanding the role of external photoevaporation is that it is only easily detected for individual systems in very high UV environments in the vicinity of O stars \citep[such as the ``proplyds'', e.g.][]{1996AJ....111.1977M, 2001AJ....122.2662O}. In such cases large cometary plumes of material are detectable in the optical, silhouetted against the ionised material of the star forming region. In the range of UV field strengths that young stars are exposed to \citep[see][for probability distributions of environmental UV fields that young stars are exposed to]{2008ApJ...675.1361F, 2020MNRAS.491..903W} these  proplyds are rare, being irradiated by about $10^5\,$G$_0$\footnote{G$_0$ is the Habing unit, which is $1.6\times 10^{-3}$ erg cm$^{-2}$ s$^{-1}$ over the wavelength range ($912$\AA$<\lambda<2400$\AA)  \cite{1968BAN....19..421H}}.  It is only in recent years that direct (i.e. not statistical) evidence for external photoevaporation has started to arise in weaker/intermediate UV fields that most young stars are exposed to. For example \cite{2016ApJ...826L..15K} observed cometary features in H$\alpha$ for discs in $\approx3000\,$G$_0$ conditions in Orion. \cite{2017MNRAS.468L.108H} also found the first evidence of a disc being externally photoevaporated in a very low UV environment ($\approx4\,$G$_0$) which is possible because the disc (IM Lup) is so large (there is emission out to 1000\,AU) and the outer regions are only weakly gravitationally bound to the star.

Our direct evidence for external photoevaporation is only for a handful of systems. We require many more observations that probe the external photoevaporation of planet-forming discs in action to confirm its importance for disc evolution and planet formation. To achieve this, we have to identify suitable observational tracers. Typically, discs are observed in the continuum and the most common line observations are in CO. 1D models of FUV-irradiated protoplanetary discs predicted that CO would actually trace a substantial part of the evaporative wind and so would be a reasonable tracer of external photoevaporation \citep{2016MNRAS.463.3616H}. However, in the multidimensional models of this process \cite{2019MNRAS.485.3895H} showed that CO is actually dissociated towards the base of the wind where deviations from a Keplerian disc are subtle, making it a poor tracer of the wind \citep[except perhaps in very weak UV environments such as IM Lup][]{2017MNRAS.468L.108H, 2018A&A...609A..47P}. Thus, CI was suggested to be a significantly better tracer of the faster wind, and hence a much better tracer with which to identify and characterise such winds. 

This paper is dedicated to using 3D models \citep[like those of][]{2019MNRAS.485.3895H} to predict and understand the observational signatures of externally photoevaporating discs in CI, to inform and enable real observing programs {that, for example, might target systems in the intermediate strength UV fields at larger separations from OB stars in Orion than the clearly evaporating proplyds.  }



 \begin{figure*}
     \hspace{-0.2cm}
     \vspace{-0.1cm}
     \includegraphics[width=16cm]{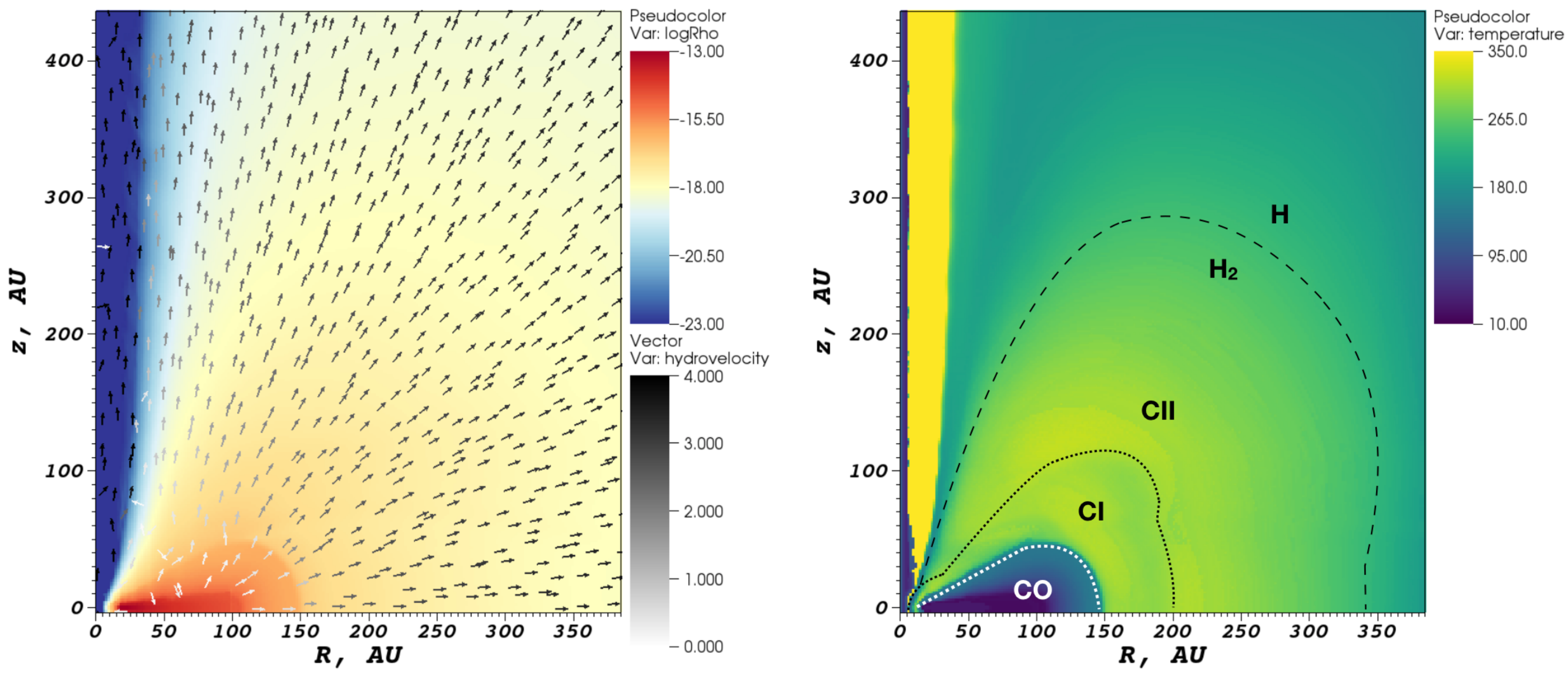}
     \caption{An example of an underlying disc photoevaporation model from which we produce our synthetic observations (this is model A from \protect\cite{2019MNRAS.485.3895H}. The left hand panel is the logarithmic density distribution (cgs, colourbar) and velocity (vectors, km/s). The right hand panel is the temperature distribution (K), with dotted contours and labels denoting where CO, CI and CII dominate in abundance over one another. The large dashed line marks the H-H$_2$ transition. }
     \label{fig:BaseModel}
 \end{figure*}
 
\begin{figure}
    \hspace{-0.8cm}
    \includegraphics[width=9.3cm]{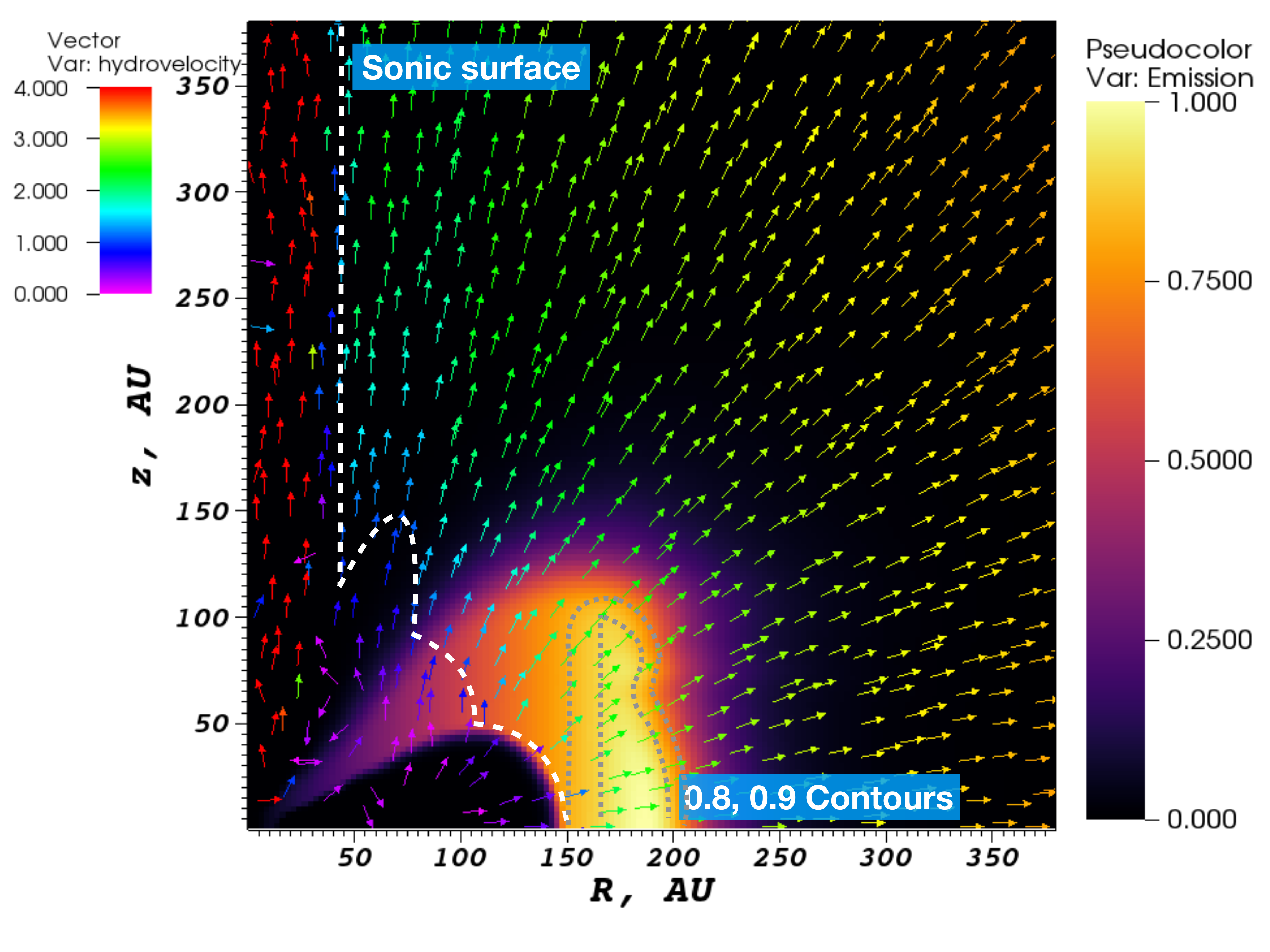}
    \caption{A map of the volume-weighted CI 3P$_1$-3P$_0$ emission for model A \protect\citep{2019MNRAS.485.3895H}, {normalized to the cell of peak emission on the grid}. Included are velocity vectors and the white dashed line denotes the sonic surface. The grey dotted lines are contours of 0.8 and 0.9 of the peak emission.  The CI emission is predominantly supersonic (a peak of $\sim2.3$\,km\,s$^{-1}$ for $c_s\approx1.1\,$km\,s$^{-1}$) and has an outwards radial component from the very outer edge and a second angled component from the disc surface. }
    \label{fig:emission}
\end{figure}

\section{Calculating synthetic ALMA observations}

We predict the observables based on 3D axisymmetric models of externally irradiated protoplanetary discs as detailed in \cite{2019MNRAS.485.3895H}. These use the \textsc{torus-3dpdr} code \citep{2019A&C....27...63H, 2015MNRAS.454.2828B} to solve for the dynamical flow structure, with thermal chemistry set by solving a photodissociation region (PDR) chemical network including 3D line cooling \citep{2012MNRAS.427.2100B} using a \textsc{healpix} scheme \citep{2005ApJ...622..759G}. 

In these dynamical models we irradiate a fixed disc that is not itself allowed to evolve. This imposed zone out to some fixed outer radius and up to some assumed vertical scale essentially acts as a boundary condition. We then iteratively update the temperature (with the 3D PDR calculation)  and the dynamical evolution until a steady state wind solution is achieved \citep[see][where we confirm our models are in a steady state]{2019MNRAS.485.3895H}. These calculations yield not just the density, temperature and velocity structure of the externally driven disc-wind, but also yield the PDR chemical composition and excitation states. In this paper we present synthetic observations from model A and B from \cite{2019MNRAS.485.3895H}. Both of these are in a 1000\,G$_0$ environment. Model A is a $10\,M_{\textrm{jup}}$, 100AU disc and model B a $20\,M_{\textrm{jup}}$, 200AU disc. The disc size can evolve quickly when external photoevaporation operates \citep{2017MNRAS.468L.108H, 2018MNRAS.475.5460H, 2018MNRAS.478.2700W, 2019MNRAS.490.5678C, 2020MNRAS.491..903W}, {and certainly would for these models which have mass loss rates $\sim10^{-6}$\,M$_\odot$\,yr$^{-1}$}, but the flow morphology and observable characteristics are similar regardless, {so long as the disc is still large enough to drive an externally driven photoevaporative wind} \citep{2019MNRAS.485.3895H}. An illustrative example of the density, temperature and velocity for model A is given in Figure \ref{fig:BaseModel}.

 These models self-consistently compute everything required to produce synthetic line observations from them. In practice we produce synthetic observations using the well established  ray-tracing framework within the \textsc{torus} code \citep{2010MNRAS.407..986R, 2019A&C....27...63H}, with abundances and level populations fed in from the dynamical-PDR model. The only quantities that we have to make further assumptions for (i.e. that are not directly computed in the PDR-dynamical model) are the microturbulent velocity (assumed to be 0.1\,km\,s$^{-1}$) and properties of the grain distribution responsible for the continuum emission (minimum/maximum grain sizes of 0.1\,$\mu$m/1\,mm and power law of distribution $q=3.3$). Note that in reality only small grains are entrained in the wind \citep[][and this is accounted for in our PDR models]{2016MNRAS.457.3593F} but we include larger grains in our synthetic observations (which would be present in the disc). Ultimately we find that the continuum contribution (which is subtracted, see below) does not qualitatively change our results, regardless of whether we include large grains. We still include the continuum so that our synthetic observations account for the process of subtracting it. Further note that in Figure \ref{fig:BaseModel} freeze out of CO onto grains has not been applied in the CO zone, but it is for our synthetic observations when the temperature is $<20\,$K and the density is $>3\times10^4$\,cm$^{-3}$.

 We simulate observations with ALMA using the CASA software \citep{2007ASPC..376..127M}. The main line that we target is the CI 3P$_1$-3P$_0$ line, which at 492.16\,GHz is in ALMA band 8. We also briefly assess the CI 3P$_2$-3P$_1$ line (which at 809.34\,GHz would require ALMA band 10) and additionally produce some synthetic CO observations for comparison. 
 
 In our datacubes we use a spectral resolution of 0.172km/s, which is finer than the anticipated sound speed ($>1\,$km\,s$^{-1}$). We assume a mm column of precipitable water vapour (pwv) of 0.472\,mm (1st octile) for all observations. This is the automatic choice for the CI 3P$_1$-3P$_0$ and CI 3P$_2$-3P$_1$ lines according to the ALMA sensitivity calculator\footnote{\url{https://almascience.eso.org/proposing/sensitivity-calculator}}, though the CO observations could be taken for higher pwv. We assume a distance of 400\,pc and direction towards Orion \citep[e.g.][]{2018arXiv181208024G}. For simplicity we just consider the case of a 90 degree position angle. 
 
 In practice for the CI 3P$_1$-3P$_0$ line, the best trade off between achieving high spatial resolution whilst recovering large angular scales in reasonable time on source is achieved by using a combination of ALMA antenna configurations C4-6 (larger baseline for higher resolution) and C4-3 (shorter baseline for larger angular scales). Our default time on source in each configuration is 120 minutes for C4-6 and 80 minutes for C4-3, which corresponds to a sensitivity of 1.13mJy and beam with semi minor and major axes of $0.070\arcsec\times0.088\arcsec$ ($28\times35$\,AU). For our CI 3P$_2$-3P$_1$ line observations we use configuration C4-5 and 40 minutes, giving a sensitivity of 0.66\,mJy and beam size of $0.072\arcsec\times0.081\arcsec$ ($29\times32$\,AU). 
 
 For our synthetic CO observations we get comparable spectral and spatial resolution using the single configuration C4-8 with 80 minutes on source, corresponding to a sensitivity of 2.35mJy and a beam size of $0.080\arcsec\times0.091\arcsec$ ($32\times36$\,AU). 

For simplicity, our CASA processing employs an automatic clean, so in practice similar quality data could be obtained for lower time on source with a detailed manual clean/reduction. However the objective here is not to predict the minimum specific required observing times (our observing times employed are not excessive) but to search for realistic signatures that are resilient to interferometric effects.  In all cases we subtract the continuum by averaging over line-free channels.

\begin{figure*}
    \hspace{-0.9cm}	
    \includegraphics[width=6.25cm]{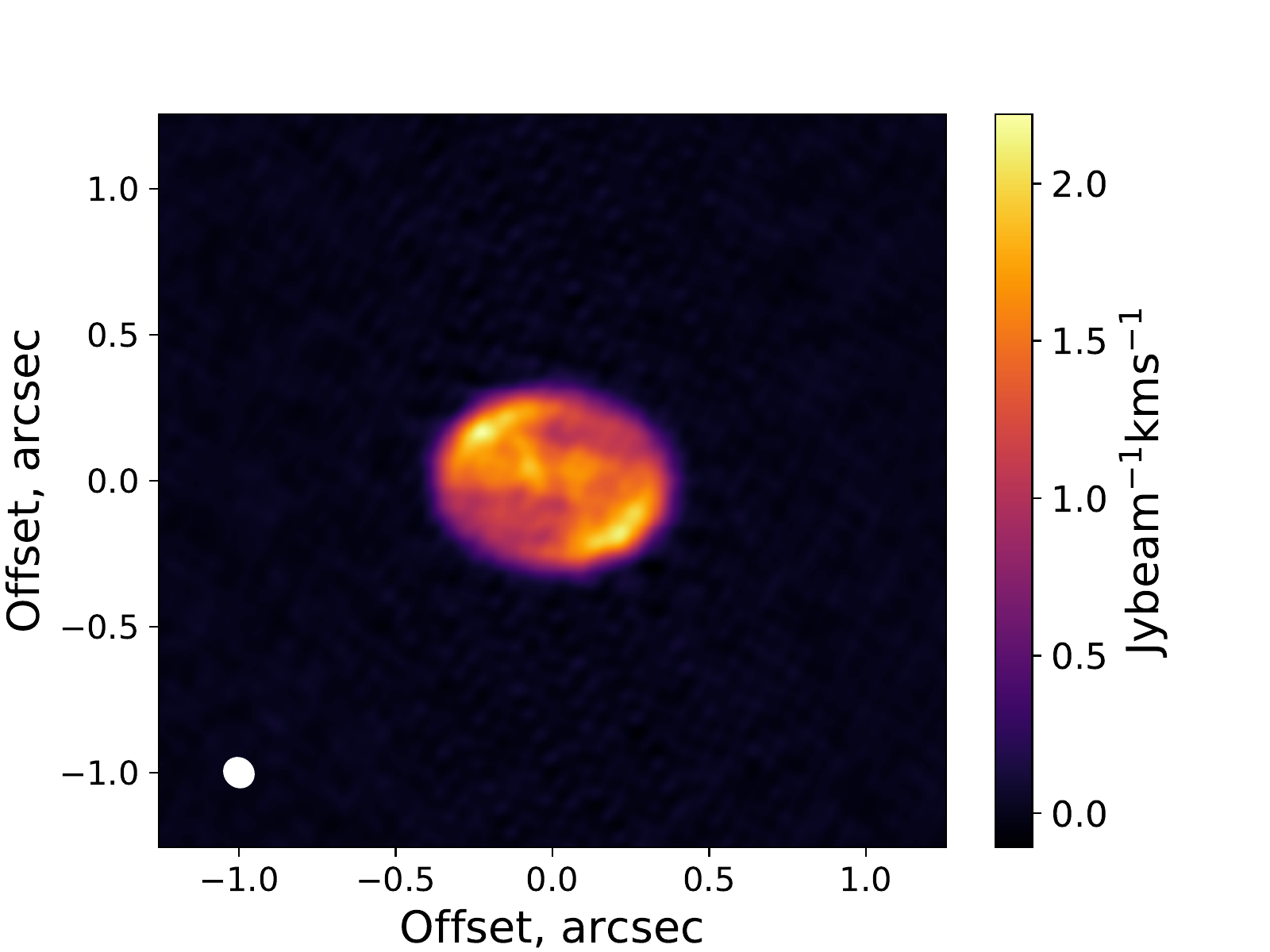}
    \hspace{-0.4cm}
	\includegraphics[width=6.25cm]{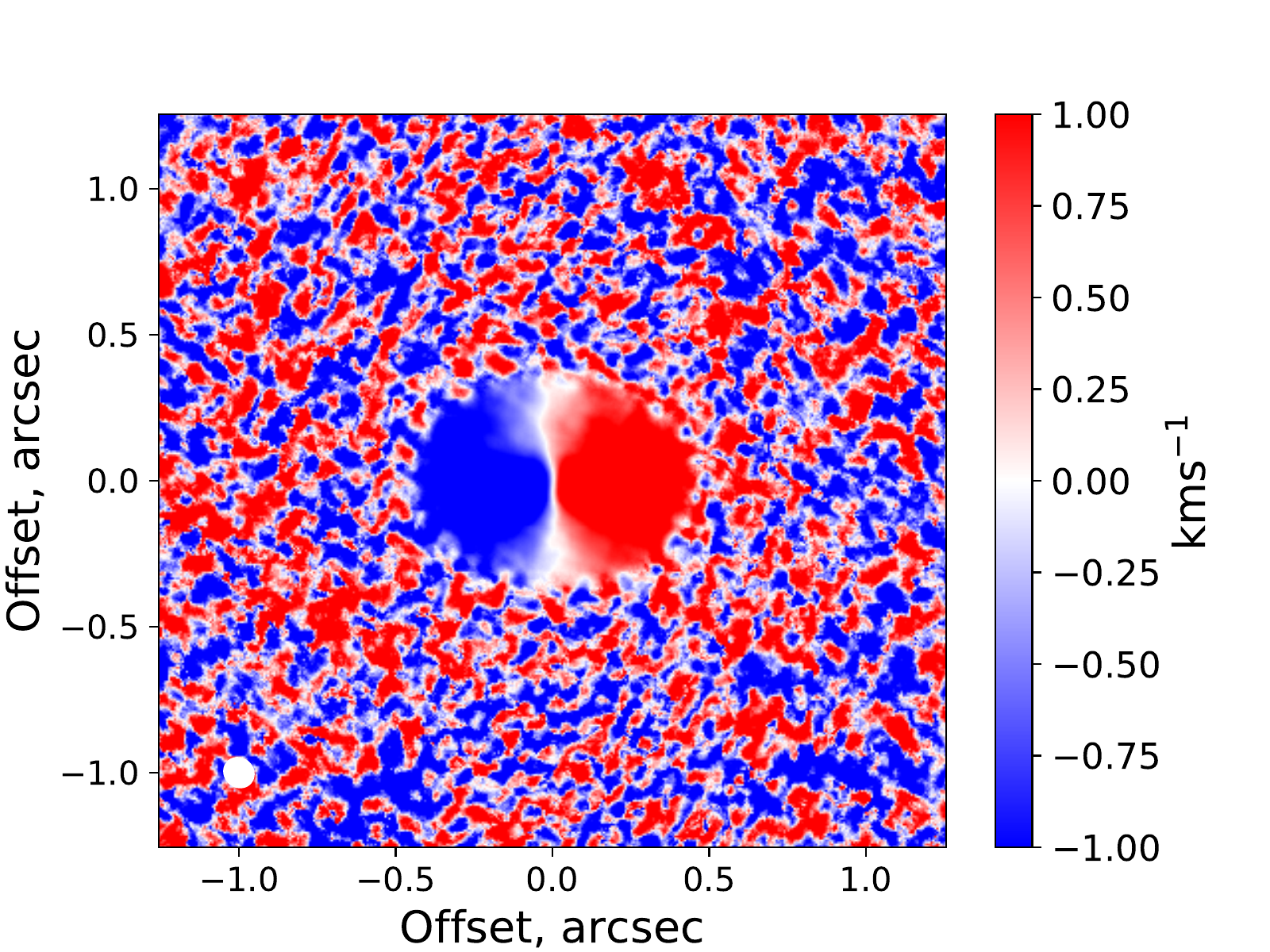}
    \hspace{-0.4cm}
	\includegraphics[width=6.25cm]{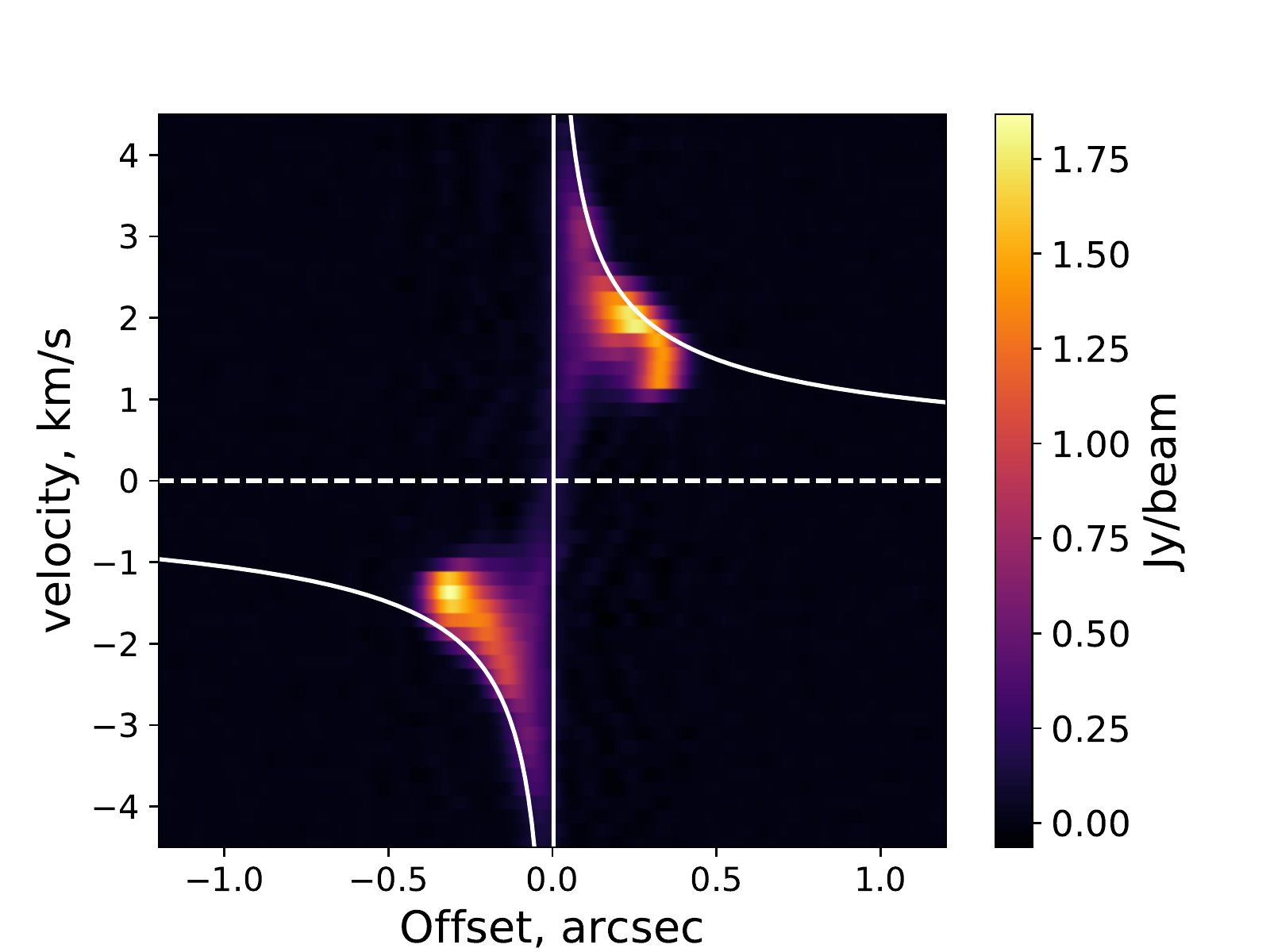}	
	
    \hspace{-0.9cm}	
    \includegraphics[width=6.25cm]{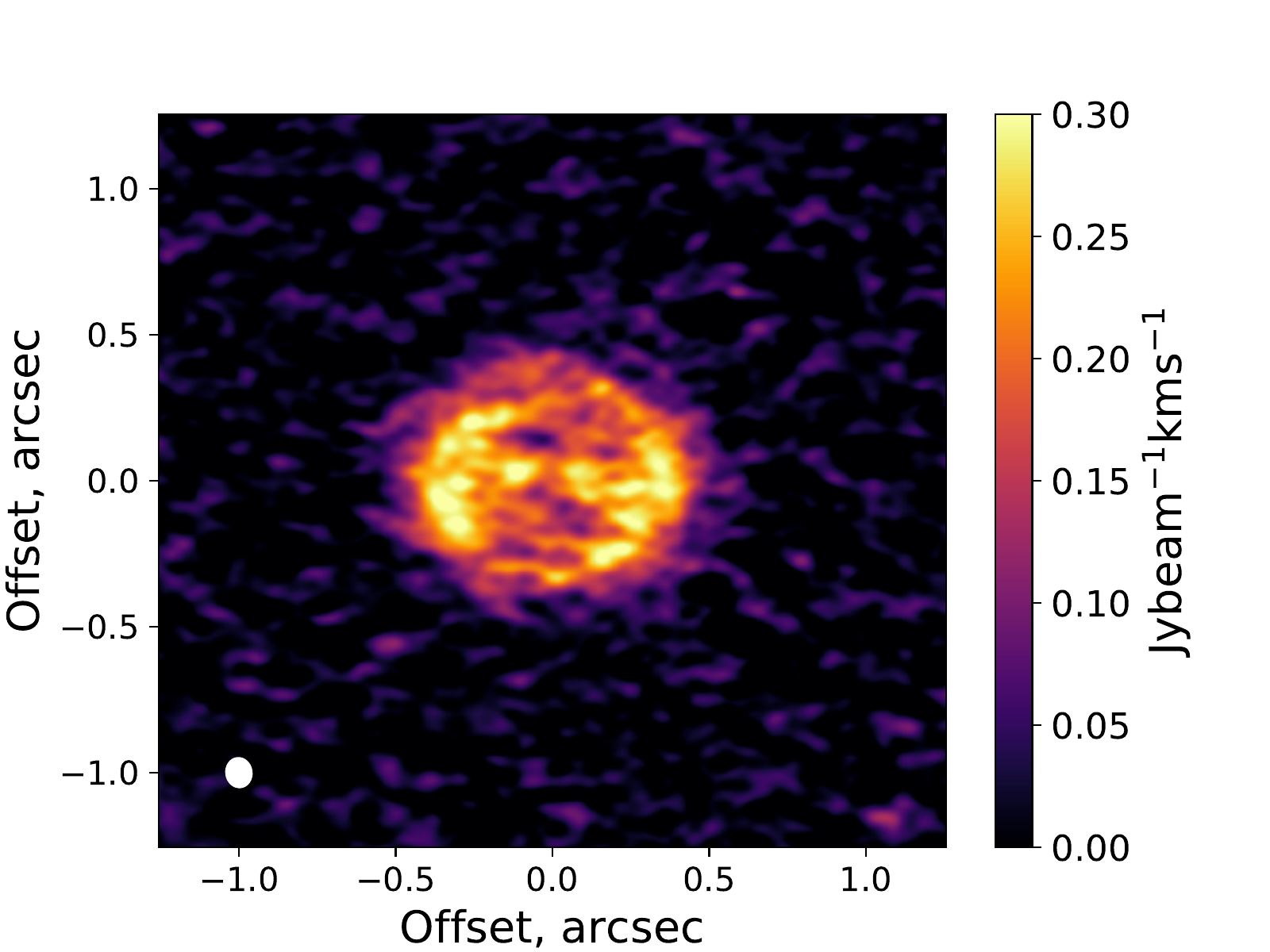}
    \hspace{-0.4cm}
	\includegraphics[width=6.25cm]{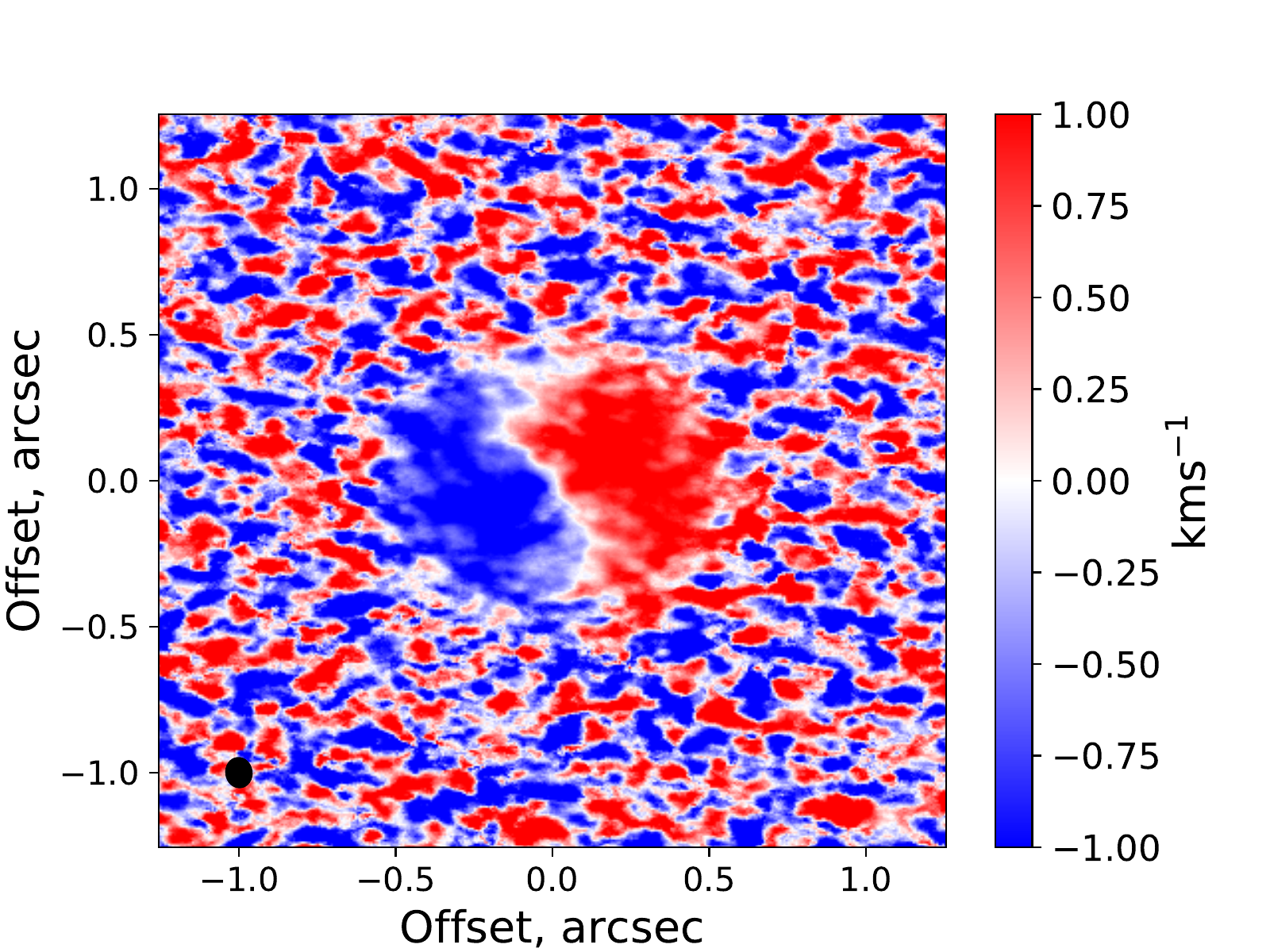}
    \hspace{-0.4cm}
	\includegraphics[width=6.25cm]{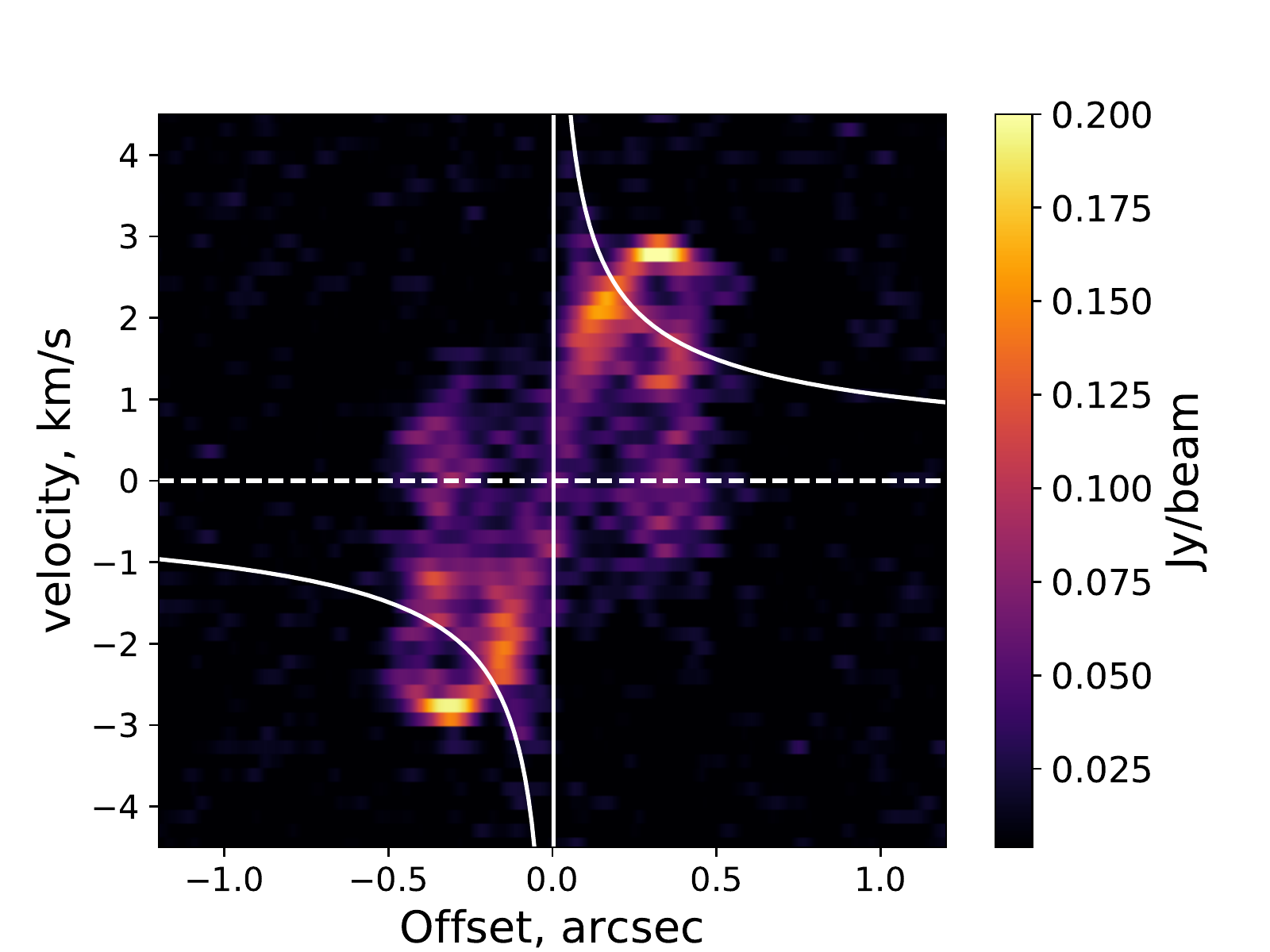}	
	\caption{45 dedgree inclination synthetic observations of model A from \protect\cite{2019MNRAS.485.3895H}, which is a 100\,AU disc in a  1000G$_0$ environment. The upper panels are CO J=1-0 and the lower panels CI 3P$_1$-3P$_0$ transitions. The left hand panels are moment 0 maps, the central moment 1 and the right hand panels are position-velocity diagrams. The curved lines overlaid on the PV diagram represent the expected boundary for Keplerian rotation.  }
	\label{fig:overview}
\end{figure*}

\subsection{The origin of emission}
\label{sec:origin}
It is important to understand from where in the wind the bulk of the CI emission is being received. The emissivity integrated over an emission line due to transitions from some upper state $u$ to lower state $l$ is
\begin{equation}
    j = \frac{1}{4\pi}n_u A_{ul} h\nu_{ul}
\end{equation}
where $n_u$ is the number density of particles in the upper state, $\nu_{ul}$ the frequency at the line centre and $A_{ul}$ the Einstein A coefficient. In Figure \ref{fig:emission} we integrate this over each model cell and normalize the resulting distribution to produce a map of from where the emission predominantly originates in model A. The grey dotted contours denote where the emission is 0.8 and 0.9 the maximum value. The white dashed line denotes the sonic surface (where the gas velocity equals the sound speed, i.e. a Mach number of 1), so the emission is predominantly from the sonic surface out to the CII front. The emission is also mostly originating from the outer parts of the zone where CI is abundant. This is the region with the largest volume in cylindrical symmetry, just before the emissivity drops near the CI-CII transition region. These characteristics are typical for our models in $\sim1000\,$G$_0$ environments. In the case of model A, the velocity of the strongest emission region is $2.3-2.5$\,km\,s$^{-1}$, which corresponds to a Mach number of $\sim2.1$.  We use the fact that the emission can crudely be thought of to come from a narrow ring in cylindrical radius to interpret our synthetic observations.

\section{Results \& Discussion}
The best tracer of an externally driven photoevaporative wind will be a function of the strength of the UV environment.

In (rare) high UV environments external photoevaporation has such a dramatic effect on the disc that the cometary shaped wind can be observed in the optical, silhouetted against the ionised HII region \citep[e.g.][]{1996AJ....111.1977M, 2001AJ....122.2662O}. $H\alpha$ can also trace winds in high-intermediate UV regimes \citep{2016ApJ...826L..15K}.

In very weak UV fields (where CO survives in the wind) CO can be used to infer photoevaporation through sub-Keplerian rotation in the flow \citep{2016MNRAS.457.3593F, 2016MNRAS.463.3616H} as in the case of IM Lup \citep{2017MNRAS.468L.108H, 2018A&A...609A..47P}. This arises because specific angular momentum ($h$) is conserved in the wind, being set by the value at the disc outer edge $h_{\rm out}$, since $h_{\rm out}\propto R_{\rm out}^{-2}$, then $v_\phi\propto R^{-1}$ in the wind, rather than the Keplerian, $v_\phi\propto R^{-1/2}$, profile. However, this sub-Keplerian rotation is a small order effect, making it difficult to detect. In addition, the majority of stars will form in intermediate strength UV environments in the 100--1000\,G$_0$ range \citep[][]{2008ApJ...675.1361F, 2020MNRAS.491..903W}. Hence, we focus our attention here on probing the kinematic signatures of winds in these common intermediate strength UV environments.  

\begin{figure*}
    \centering
    \includegraphics[width=17cm]{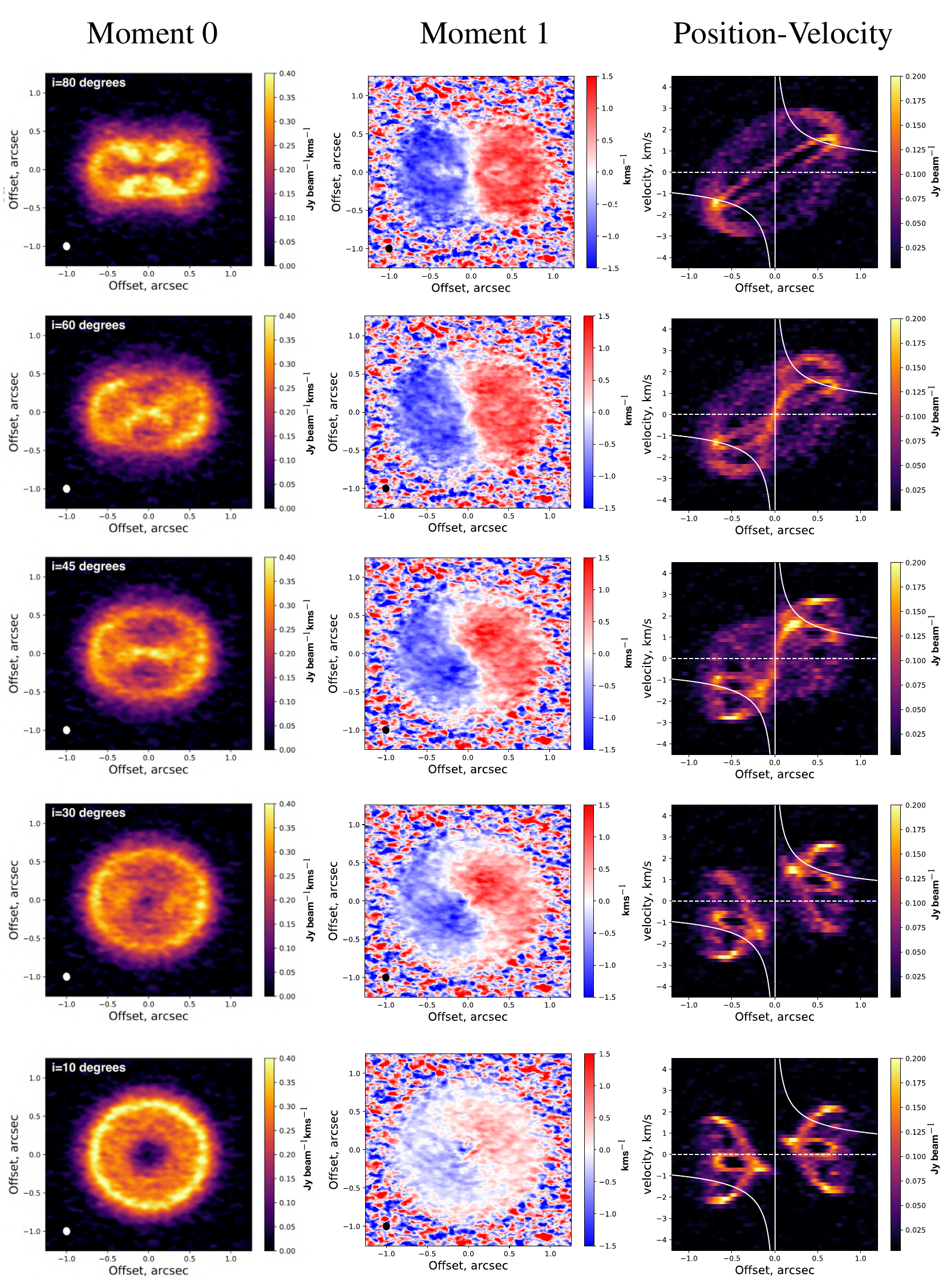}
    \caption{Moment 0 (left), moment 1 (central) and PV diagrams (right) for model B from \protect\cite{2019MNRAS.485.3895H}. This model is a 200\,AU disc in a 1000\,G$_0$ UV environment. The inclination is given in the upper left of each moment 0 map (where face-on inclination is zero). Offset ticks are in intervals of $0.5\arcsec$ and velocity ticks are at intervals of 1km\,s$^{-1}$}
    \label{fig:inclination}
\end{figure*}

\subsection{Anatomy of an evaporating discs in CO and CI}
At intermediate UV field strengths, of order 1000\,G$_0$, archetypal CO and CI moment 0, 1 maps and a position-velocity (PV) diagram at an inclination of 45 degrees are shown in Figure \ref{fig:overview}. Note that the PV diagrams are constructed from a slice across the semi-major axis of the disc. {For discs, PV diagrams to capture a Keplerian rotation curve are usually made from a cut across the whole semi-major axis that is as thin as possible, so that there is reasonable signal to noise but not too much averaging over the disc. We hence also use a thin cut across the disc for our PV diagrams. }

\begin{figure*}
	\includegraphics[width=8.5cm]{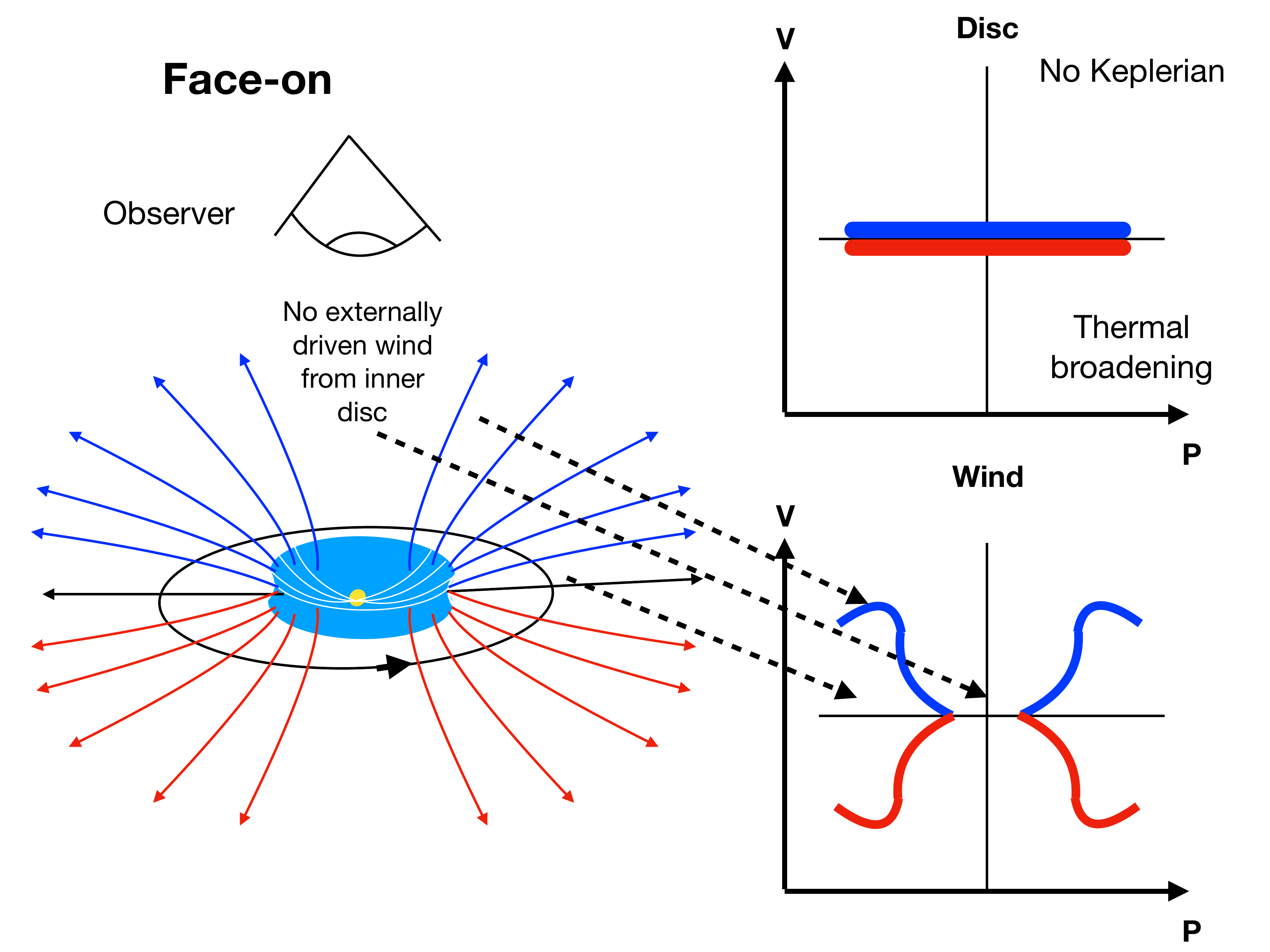}
	\includegraphics[width=8.5cm]{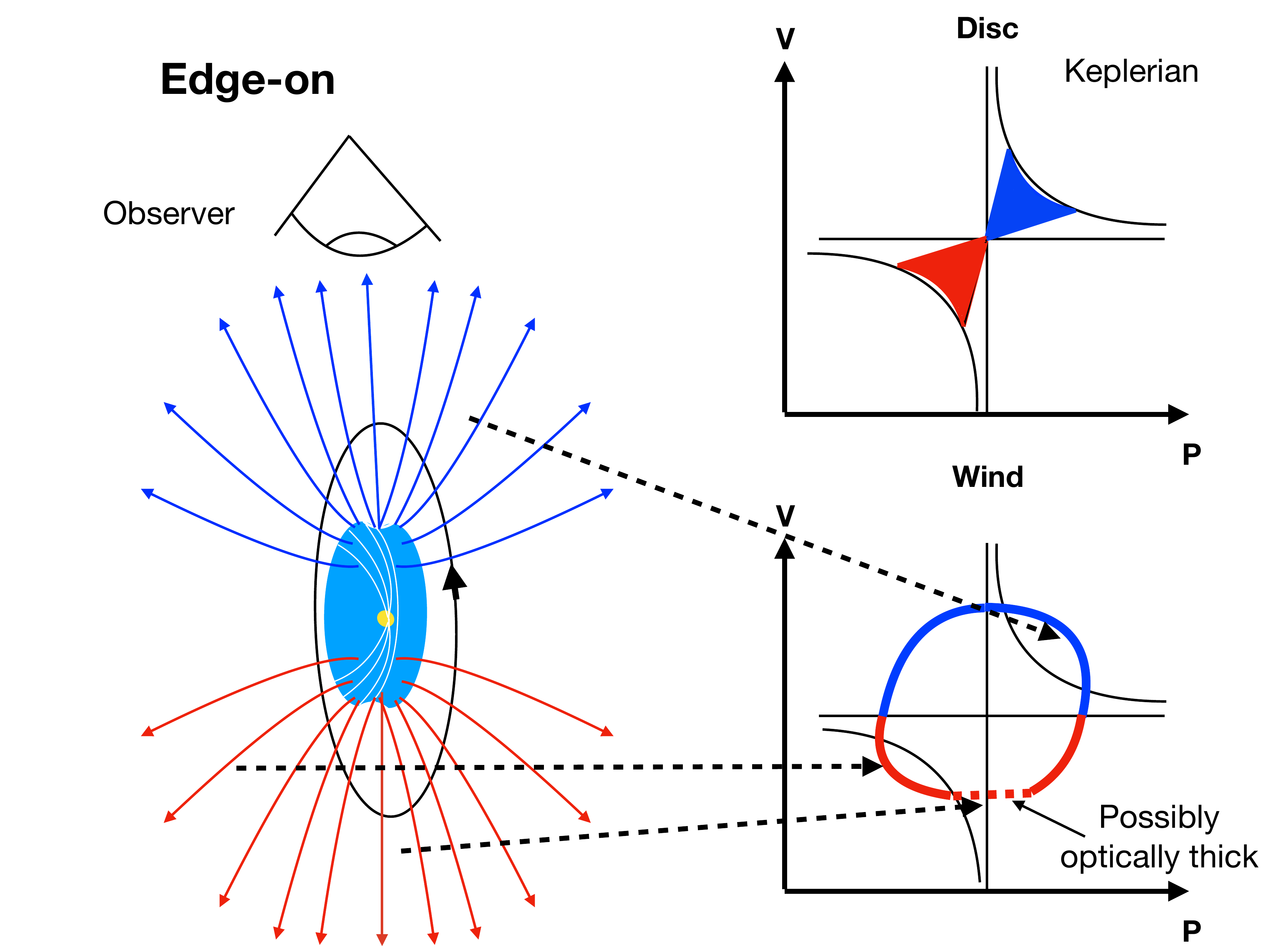}
	\caption{An illustration of the structure of PV diagrams for face-on (left) and edge-on (right) photoevaporating discs. For face-on discs there is no contribution from the Keplerian rotation, the disc itself only contributes through the thermal/turbulent broadened line at the rest frequency. The wind is not launched from the inner disc, so the inner part of the PV diagram is associated with low velocity emission (internal photoevaporative winds are too ionised to be observed in CI 3P$_1$-3P$_0$). For a face-on disc, parts of the wind driven from disc surface layers contribute to the highest velocity emission, but the main bulk of the wind driven from the disc outer edge is not observed. The concatenation of the disc and wind featurers of this face-on schematic explain the morphology of the near face-on disc in the bottom row of Figure \ref{fig:inclination}. In the edge-on case the disc contributes a Keplerian profile. Due to the flaring nature of the wind from the disc outer edge, the wind contributes a ring structure in PV space. Again, concatenation of the disc and wind morphologies gives the approximate structure in the PV diagram of the upper right panel of Figure \ref{fig:inclination}.}
	\label{fig:PVschematic}
\end{figure*}

\begin{figure}
	\includegraphics[width=9.1cm]{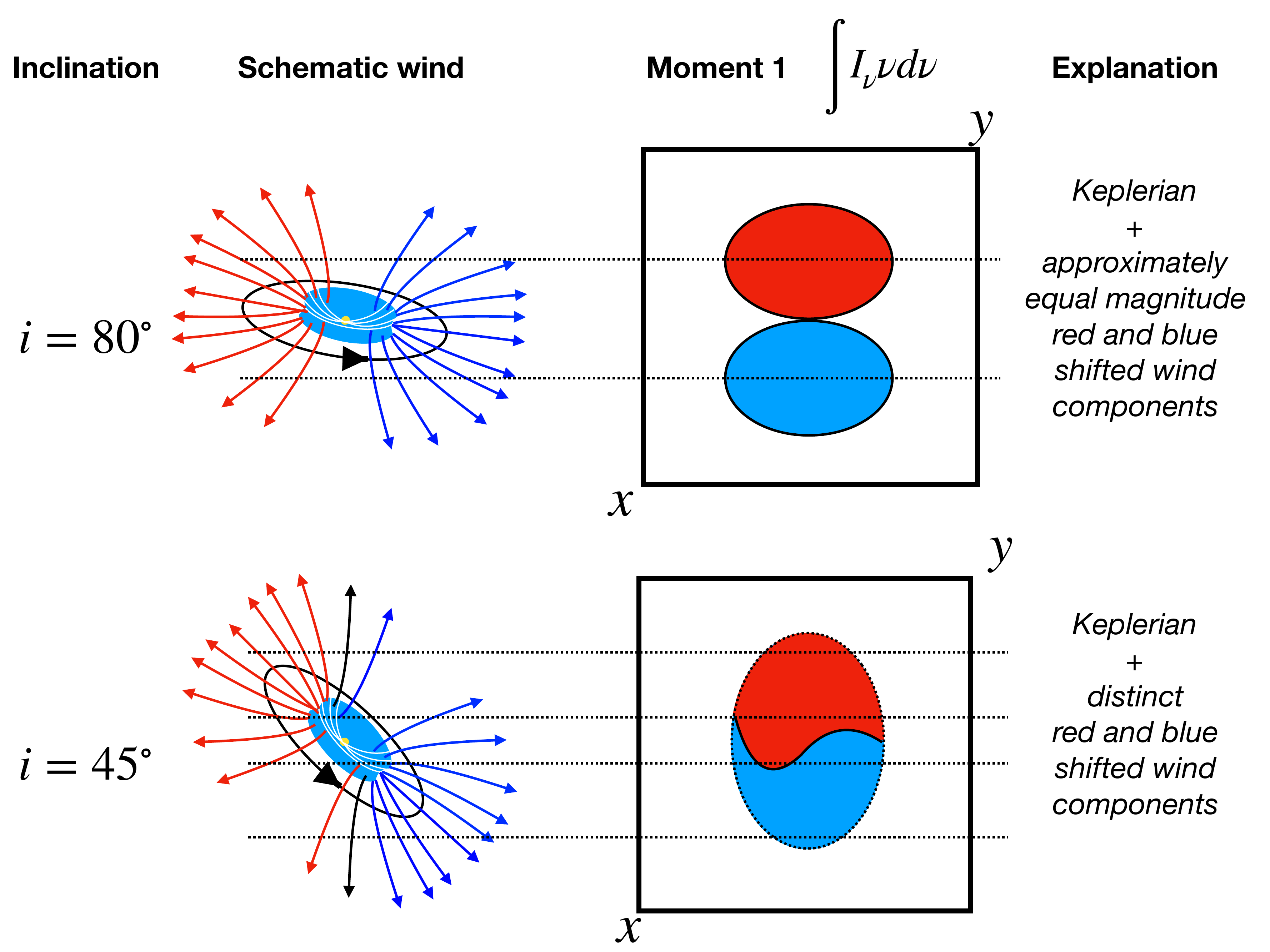}
	\caption{Approximately edge-on discs (upper panels) look Keplerian in CI moment 1 maps because the wind components along the line of sight are roughly equal and opposite. However once the disc becomes more inclined we observe different wind components along the line of sight which distorts the moment 1 map.    }
    \label{fig:mom1schematic}
\end{figure}

These models are generated from model A \citep{2019MNRAS.485.3895H} that was summarised in Figures \ref{fig:BaseModel} and \ref{fig:emission}. The imposed disc for model A had 10\,M$_{\textrm{jup}}$ of material within 100\,AU, irradiated isotropically by a 1000G$_0$ UV field.  In Figure \ref{fig:inclination} we also show the moment 0, 1 maps and PV diagrams for different inclinations in the case of model B \citep{2019MNRAS.485.3895H}, which imposes 20\,M$_{\textrm{jup}}$ within a 200AU disc in a {1000\,G$_0$} UV field. The key points to take from Figures \ref{fig:overview} and \ref{fig:inclination} are 
\begin{enumerate}
    \item Our synthetic observations in CO lines are qualitatively just like purely rotational discs \citep[though a small deviation from Keplerian rotation in the subsonic wind could be measured in principle][]{2016MNRAS.457.3593F, 2016MNRAS.463.3616H}. That is, the moment 1 map has a red-blue asymmetry about the rotation axis, characteristic of Keplerian rotation profile.  Furthermore the emission in PV space is confined to the upper right and lower left quadrants at speeds no more than Keplerian. This is because CO is photodissociated except for in the slow inner part of the wind. \\
    \item Unlike CO,  CI traces a faster component of the wind as indicated in Figure~\ref{fig:emission}. The CI PV diagram is grossly distinct from that of a Keplerian disc. The PV diagram will take on a slightly different qualitative morphology depending on the inclination (Figure \ref{fig:inclination}). However, there are two features of external photoevaporation that are always present regardless of inclination, which are 
\begin{enumerate}
    \item Emission at greater than Keplerian speeds (as a result of the wind speed). 
    \item Emission from quadrants in PV space where there is none for a Keplerian disc (the upper left and lower right). This arises because we detect emission from the radial part of the wind on both the near an far sides of the disc (see Section~\ref{sec:rings}). 
\end{enumerate}
\begin{figure}
    \hspace{-0.4cm}
    \includegraphics[width=9.4cm]{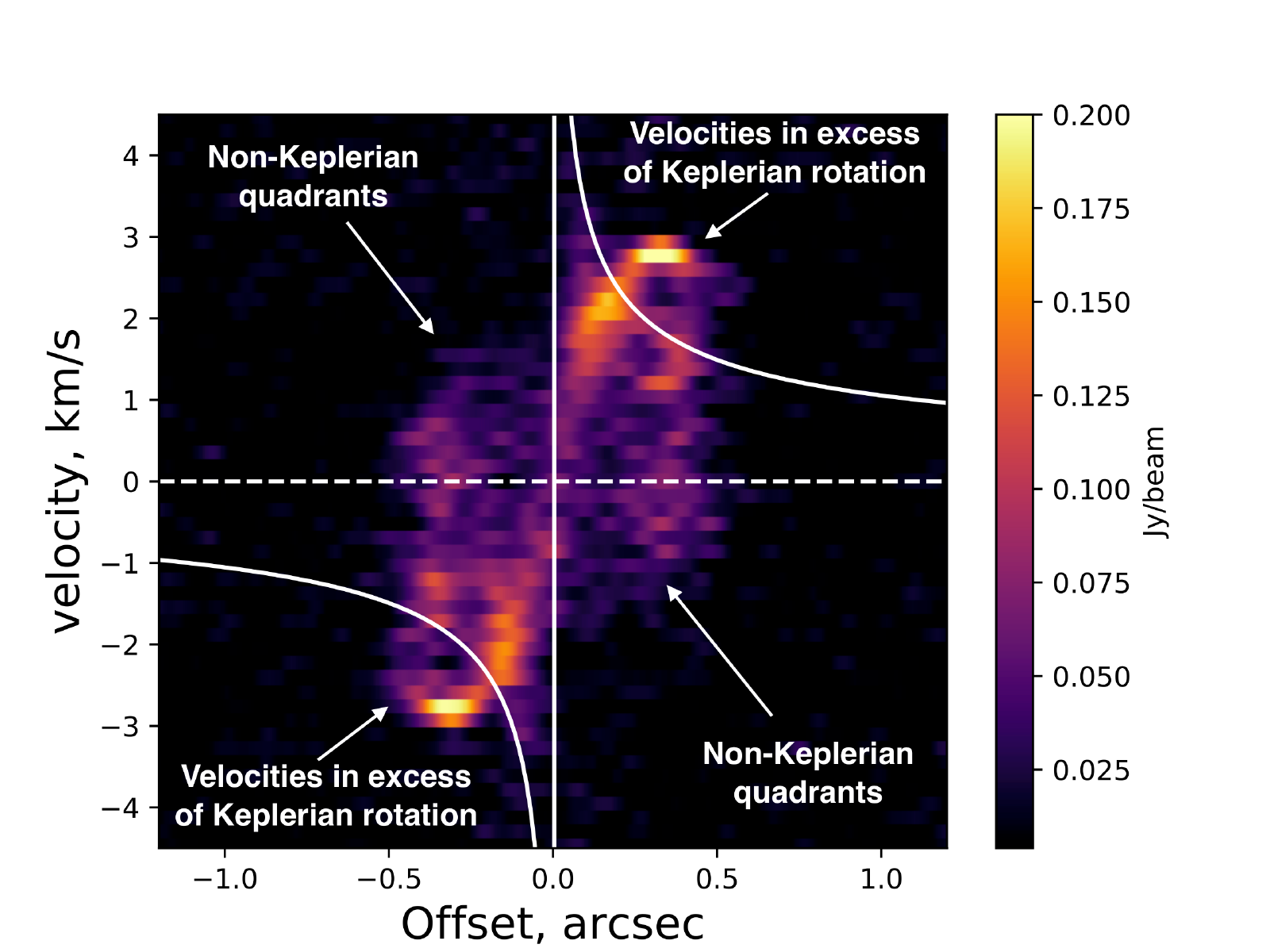}
    \caption{A PV diagram of model A from \protect\cite{2019MNRAS.485.3895H}, that illustrates the deviations from Keplerian behaviour (which would be bound by the solid white line) that are characteristic of external photoevaporation in intermediate strength UV environments. That is, velocities in excess of Keplerian rotation in the upper right, lower left quadrants and non-zero emission in the upper left, lower right quadrants. {Note that the velocities in excess of Keplerian rotation stem from radial components of the flow, not super-Keplerian rotation.} The synthetic observation is of a disc with 45 degree inclination.  }
    \label{fig:annotated}
\end{figure}

    These features are illustrated on PV diagram in Figure \ref{fig:annotated}.\\
    \item As mentioned above, CI PV diagrams for evaporating discs exhibit a range of morphology depending on inclination, including arcs and loops. A schematic of discs and their PV diagrams for the extreme cases of face-on and edge-on is given in Figure \ref{fig:PVschematic} to help with qualitative interpretation of how these different features arise. We will also quantify this with a simple model in section \ref{sec:rings}. At near face-on inclinations there is a remarkable ")(" morphology. {This arises because the cut for the PV diagram covers the entire length of the semi-major axis, but only a finite width along the semi-minor axis of the disc}. At the edges of the cut across the disc we see the vertical component of the wind $\pm v_z$ from the near and far side of the disc. As we move along the cut to the inner parts of the disc the  wind gets weaker (see Figures \ref{fig:BaseModel}/\ref{fig:emission}), with no wind at all being driven from the inner regions by the external radiation field. {If the width of the cut for the PV diagram were as thick as (or thicker than) the disc the maximum velocity of the ")(" would be present at all offsets as we would always see the peak $\pm v_z$.} \\
    \item The moment 1 map of an externally evaporating disc can also exhibit winding at intermediate inclinations (see the 45 and 60 degree inclination models in Figure \ref{fig:inclination}). A schematic of why this is the case is shown in Figure \ref{fig:mom1schematic}. {In short,} at near edge-on inclinations the wind velocity is symmetric and so the red and blue shifted components cancel out when we integrate through the cube to produce the moment map (this is also the case for face-on discs, where there is also no Keplerian rotation to observe). However at intermediate inclinations we observe different components of the wind along the line of sight.  Note that twists in moment 1 maps can appear due to other mechanisms such as fast radial flows \citep{2014ApJ...782...62R} or warps \citep{Casassus2015} through transition disc holes, though these are distinct because they are at smaller scales in the disc and are present in CO. 
\end{enumerate}
We hence immediately have expected observational signatures in PV diagrams and moment 1 maps that are predicted from external photoevaporation models that would not arise for a normal disc and would not manifest in CO observations. Furthermore, these characteristics of atomic carbon in the wind are distinct from inner photoevaporative winds driven by the host star because the gas is in the external wind is much cooler. Having these predicted signatures is important in its own right, since we have previously not known what to search for in intermediate UV regimes to find external photoevaporation, however it is also critical to be able to further interpret and characterise the wind, to which we now turn our focus.

\subsection{Interpreting the CI morphology in PV diagrams}
\label{ref:interpMain}
Usually when interpreting the the kinematics of discs, the dominant velocity is azimuthal rotation $v_\phi$ (which can in the first instance be approximated as Keplerian). Other kinematic components are small order effects that usually require sophisticated azimuthal averaging techniques to be revealed \citep[e.g.][]{2018ApJ...860L..13P, 2019Natur.574..378T}. 

In the case of external photoevaporation the situation is more complex since any point in the wind can also have comparable (or even dominant) velocity in the radial ($v_R$) or vertical ($v_z$) directions. We have already shown in Figures \ref{fig:overview} and \ref{fig:inclination} that this results in twists in moment 1 maps and loops and arcs in PV diagrams. Here we interpret the CI morphology in PV diagrams and assess what physical parameters can be extracted from them.

\subsubsection{Optical depth of the line}
\label{sec:opticalDepth}

CO is well known to be optically thick in discs and hence only traces surface layers. To interpret observations of external disc photoevaporation in CI 3P$_1$-3P$_0$ it is useful to anticipate the optical depth of the line to understand which parts of the disc/wind we are tracing emission from. We do so by moving along cells of the model in the disc mid-plane, and from these tracing perpendicular rays in the $\pm z$ directions along which we calculate the optical depth. We also do this for rays along any given observer viewing angle. This ray tracing is done for each velocity channel in our datacubes. 

The near-side wind is always optically thin regardless of the inclination or velocity. However, the wind from the far side of the disc can be partially optically thick (due to dust absorption) and this becomes stronger for more edge-on discs (where there is a larger column from the disc itself along the line of sight). For example in model B there is a peak optical depth of $\sim2.5$ along the maximum column (offset of zero in our PV diagrams) and this is lower for increasing offset. Optical depth may locally reduce the flux from the far-side wind at low offset (see right hand schematic in Figure \ref{fig:PVschematic}) but is not anticipated to completely suppress the far-side wind.

\subsubsection{The kinematics of rings of emission}
\label{sec:rings}
We lay the foundation for interpretation of the complicated CI PV diagrams of evaporating discs by considering the simplest case of a ring of optically thin material, as motivated by Figure~\ref{fig:emission}. We systematically vary the $R, Z$ and $\phi$ components of velocity at each point on the ring to assess how they manifest in PV diagrams. 

In appendix \ref{sec:derive} we derive a parametric form for the features that result in PV space in this simple optically thin ring model which has radius $a$, wherein for position coordinate 
\begin{equation}
\textrm{pos} = a\cos(E)
\end{equation}
the corresponding velocities are
\begin{equation}
\textrm{vel} = v_R\sin i \sin(E)+v_\phi\sin i \cos (E)\pm v_Z\cos i
\end{equation}
Here $v_R, v_\phi, v_z$ are the radial, azimuthal and vertical components of velocity, $i$ is the disc's inclination and $E$ is a parameter that runs between 0 and $2\pi$ (formally the eccentric anomaly).  In section \ref{sec:infer} we examine how this can be used as a diagnostic of real observations.  

\begin{figure*}
	\hspace{-1.1cm}
	\includegraphics[height=11.5cm]{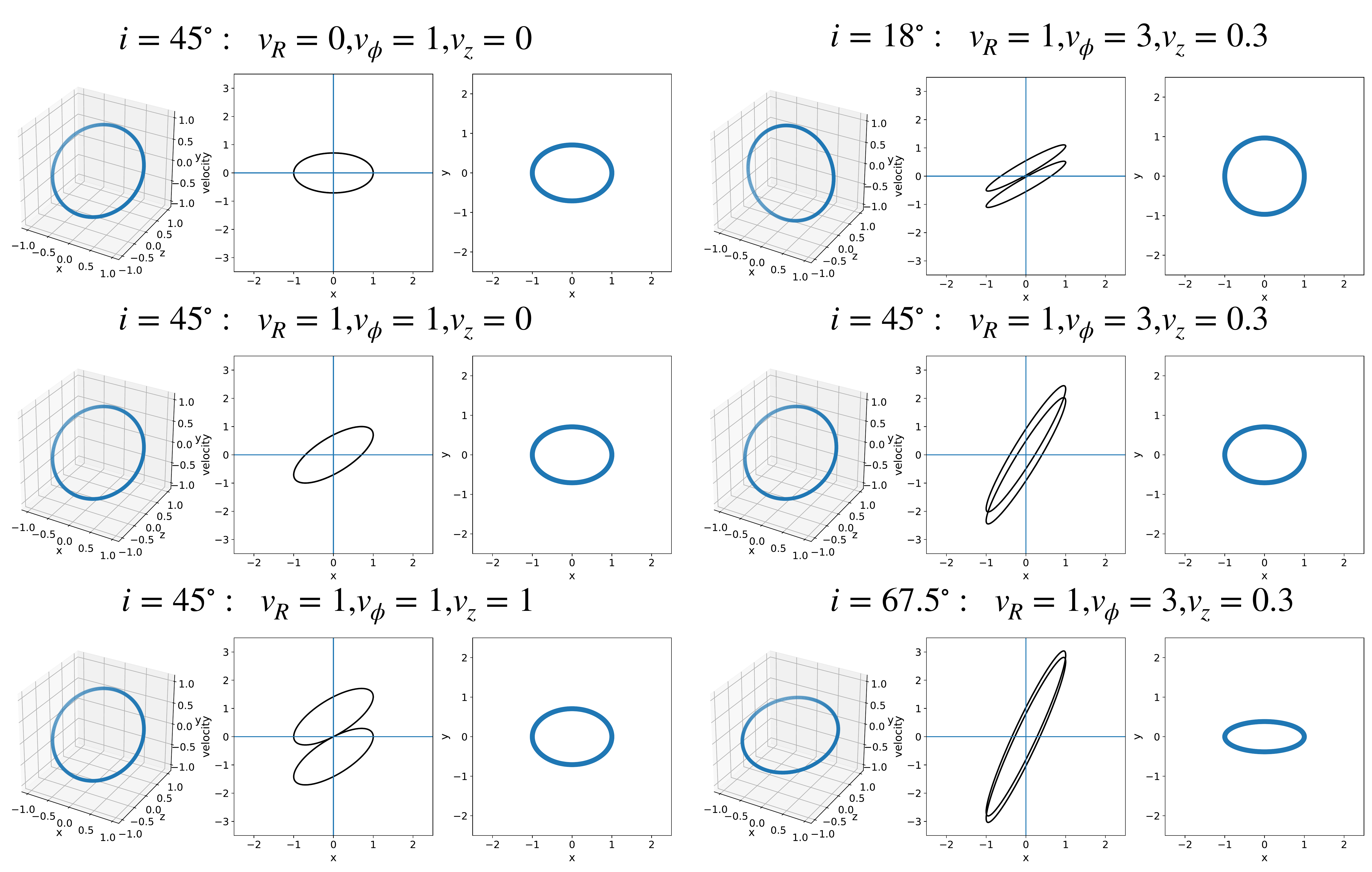}
	\caption{ A selection of our simple ring models. In each set of 3 panels we show the 3D ring structure (left), PV diagram (middle) and projected structure (right).    The left hand column is all viewed at 45\,degrees inclination. The upper left collection of panels have purely rotational velocity. The left-middle panels have rotation plus an equal magnitude outward radial component, which transforms the line of the first panel into an ellipse. The lower left panels are equal parts rotational, radial and vertical (in both $\pm z$), which results in two ellipses offset by $\pm v_z$. The right hand sets of panels all have dimensionless rotational, radial and vertical velocities of 3, 1 and 0.3 respectively, differing in their inclinations from $\pi/3$, $\pi/4$ to $\pi/8$ from top to bottom respectively.    }
	\label{fig:simpleRing}
\end{figure*}

Figure \ref{fig:simpleRing} summarises the main phenomenological changes introduced through different kinematic components of this ring model. Each set of 3 panels in this figure shows the 3D ring structure on a Cartesian grid (left panel) the PV diagram (central panel) and the 2D projected ($x-y$) ring structure. 

The upper left hand set of panels  of Figure \ref{fig:simpleRing} are for a purely rotational ($v_\phi$) ring inclined by 45\,degrees. The PV diagram is of the familiar form for a Keplerian disc, cutting from the lower left quadrant through the origin to the upper right quadrant. As the inclination changes the angle of this line would change (being shallower for a more face-on disc).

The left-central set of panels then adds an outwards radial velocity ($v_R$) component to the ring of the same magnitude as the rotational component. This has the effect of transforming the PV line into an ellipse with a semi-major axis the same as the line in the rotation-only case. A rotational-radial flow immediately deviates strongly from a rotational-only flow, with emission appearing in the top-left and lower-right quadrants of the PV diagram, which otherwise exhibit no emission. Conceptually, in the rotational only case the sign of the velocity is not sensitive to the near or far side of the ring being observed, only on the half of the disc being observed. Conversely when the flow has an outwards radial component, at all positions the velocity from the near and far side of the ring have opposite sign, which is why an ellipse results. 

Parts of external photoevaporative winds emanating from the disc surface can also have a strong vertical component (see left hand panel of Figure \ref{fig:BaseModel} and Figure \ref{fig:emission}). In the lower right set of panels of Figure \ref{fig:simpleRing} we introduce a $\pm z$ component of velocity ($v_z$) to the ring of emission (which still has rotational and radial components of the same magnitude) and demonstrate that this results in two rings offset by $2v_z$. 

This simple model of a kinematic ring predicts that emission from rotating rings with a radial component will be an ellipse, and that if there is a vertical components there will be multiple (potentially concentric) ellipses. Multiple rings appear in our synthetic PV diagrams (dependent on viewing angle) which, according to this simple ring model, are explained by radial components of the wind, as well seeing near and far-side vertical components of the wind.

Our discussion so far has considered the radial, rotational and vertical components of velocity with equal magnitude. Their actual ratios will be somewhat dependent on the nature of the disc and incident UV field, but by inspection of model A from \cite{2019MNRAS.485.3895H} (100\,AU disc, 1000\,G$_0$ UV field) at $\sim200$\,AU in the main region of emission emanating from the very disc outer edge we find the relative strengths of  $v_\phi, v_R, v_z $ are approximately $1, 0.3, 0.1$ normalised to the rotational velocity. In the right hand panels of Figure \ref{fig:simpleRing}  we show rings with these speeds (scaled up to $v_\phi=3$). The upper, central and lower sets of panels are for inclinations of $\pi/3, \pi/4$ and $\pi/8$ respectively. Although detailed features, which arise from optical depth effects, are clearly missing from these simple ring models that appear in our synthetic observations, they demonstrate that concentric ellipses are a natural product of a rotating ring of material that also has a radial velocity component (making the ellipse) and vertical component (offsetting a second ellipse). Additional complexities arise due to variations in the radial and vertical structure of the flow, and the optical depth (section \ref{sec:opticalDepth}). 

\subsubsection{Constraining the physical parameters of real evaporating discs}
\label{sec:infer}
So far we have discussed qualitative signatures of moment 1 maps and PV diagrams that are indicative of an external photoevaporative wind. We now turn our attention to inferring quantitative parameters of the wind itself. 

The most accessible way to infer quantitative information is through the PV diagrams. It is possible to use the ellipse scheme of section \ref{sec:rings} to interpret real observations of evaporating discs, in particular if the inclination is known from, e.g. CO or continuum observations. One can then determine for what velocities the best match to the PV features are obtained. 

\begin{figure}
    \centering
    \includegraphics[width=9.4cm]{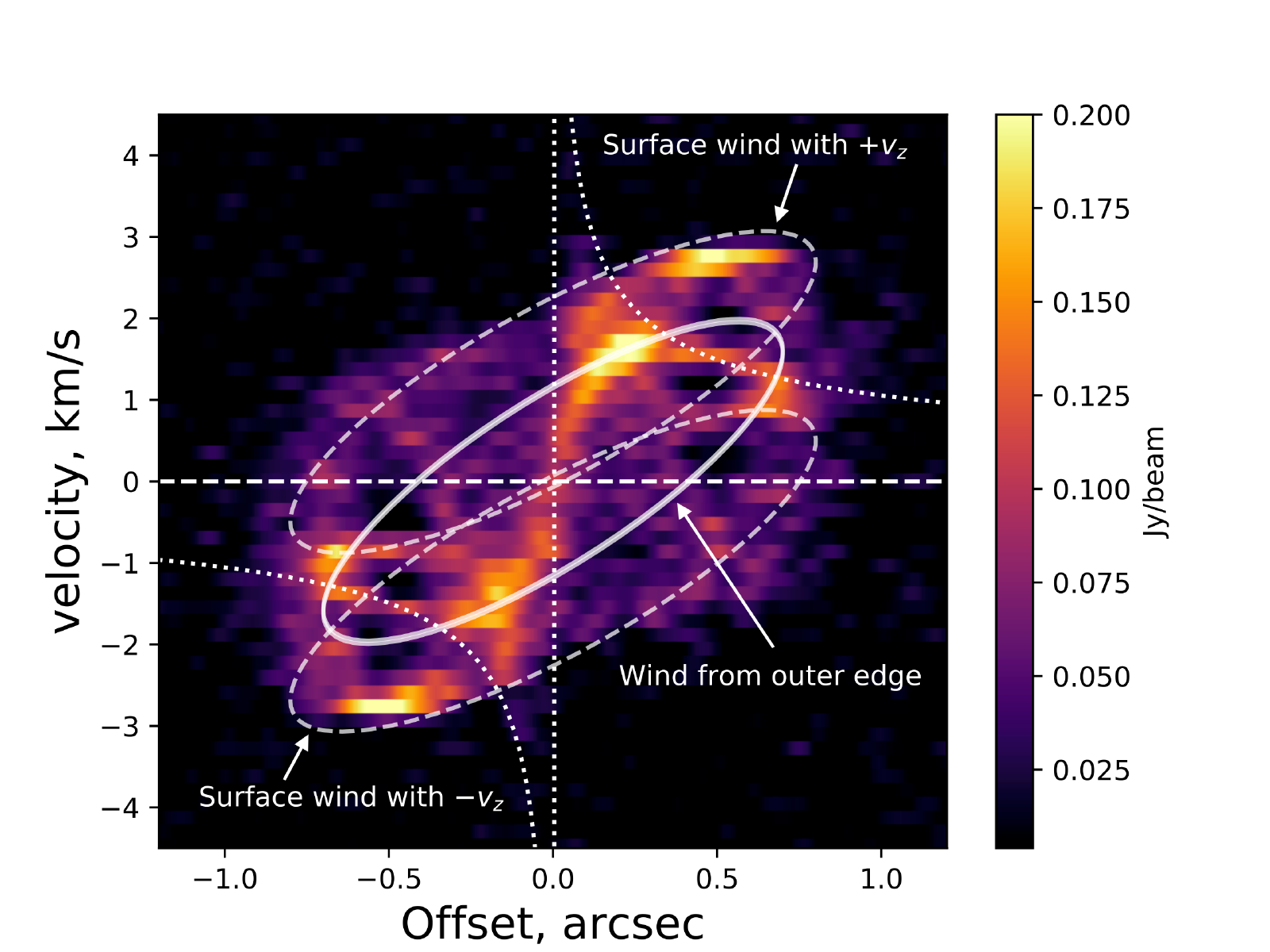}
    \caption{Our simple optically thin ring of emission model overplotted on the model B PV diagram at 45 degree inclination. Here the ring properties were informed by the underlying model, with one set representing the flow from the very outer edge of the disc ({which has negligible $\pm v_z$, solid line}) and the other representing the flow from the disc surface ({dashed lines corresponding to $\pm v_z$}). In principle rings could be fitted to a real PV diagram. {The parameters used for the ring model are $v_R, v_\phi, v_z = $ [1.65, 2.25, 0.03] and [1.65, 2.25, 1.55]\,km\,s$^{-1}$ for the mid-plane and surface wind respectively.}  }
    \label{fig:fitPV}
\end{figure}

However, in practice it is difficult to determine how many ellipse components need to be fit to a real PV diagram. For example, consider fitting the middle-right panel of Figure \ref{fig:inclination} (model B, inclined 45 degrees). In model B we \textit{know} that there is a dominant kinematic component from the disc-edge and an angled component of the wind from the disc surface (it is very similar to the map of volume-weighted emission for model A in Figure \ref{fig:emission}). In Figure \ref{fig:fitPV} we overplot ellipses of emission with appropriate velocities and sizes of these two zones ($v_R, v_\phi, v_z = $ [1.65, 2.25, 0.03] and [1.65, 2.25, 1.55]\,km\,s$^{-1}$ for the mid-plane and surface wind respectively) onto the 45 degree PV-diagram. The ellipses do a good job of corresponding to the main features, but the form is not what one would necessarily choose to fit without knowledge of the underlying flow. The flow is complicated and could be optically thick in places, so the ellipses do not correspond to loops of uniform intensity. This is alleviated to some degree by the fact that our models usually have a similar flow structure, but for any real system this may not be the case, particularly if the UV field deviates significantly from being isotropic. {We therefore retain the ellipse scheme only as a means of interpreting the broad kinematic morphology here. }

It is worth noting that at zero offset there is zero azimuthal contribution, with the velocity being set purely by the radial and vertical components of velocity. If there are two ellipses separated by $2v_z$ we can then extract the radial and vertical components of the wind. 

Overall then, although the rings of emission are extremely useful for providing a basis for understanding the general features we observe and can in principle be used to infer the flow kinematics, using them in practice to infer $v_R, v_\theta, v_\phi$ needs careful consideration and appreciation of the uncertainty/possible degeneracy if applied in practice. We reiterate that optical depth becomes more important for near edge-on discs and even then at small offset, so should not be a limiting factor for using this emission ring scheme in practice (section \ref{sec:opticalDepth}). 

 Since CI survives from approximately the sonic surface out to speeds of about Mach 2, PV diagrams can also be used to place weak constraints on the thermal properties of the CI region of externally photoevaporating discs. We specify weak since the temperature would scale with the square of the sound speed.  For example for model B a velocity of 2.5\,km\,s$^{-1}$ would place an upper limit on the temperature of $\sim1500\,$K, whereas in the model the temperature of the CI zone is $\sim$300\,K.

\subsection{CI 3P$_2$-3P$_1$ Band 10 observations }
Our main focus has been the band 8 CI 3P$_1$-3P$_0$ line, however the 3P$_2$-3P$_1$ line at 809\,GHz could also be observed in ALMA band 10. Complex atmospheric absorption means that band 10 requires extremely good observing conditions. Nevertheless, we now briefly assess the insights and opportunities that band 10 observations would provide.

\begin{figure}
    \centering
    \includegraphics[width=9.4cm]{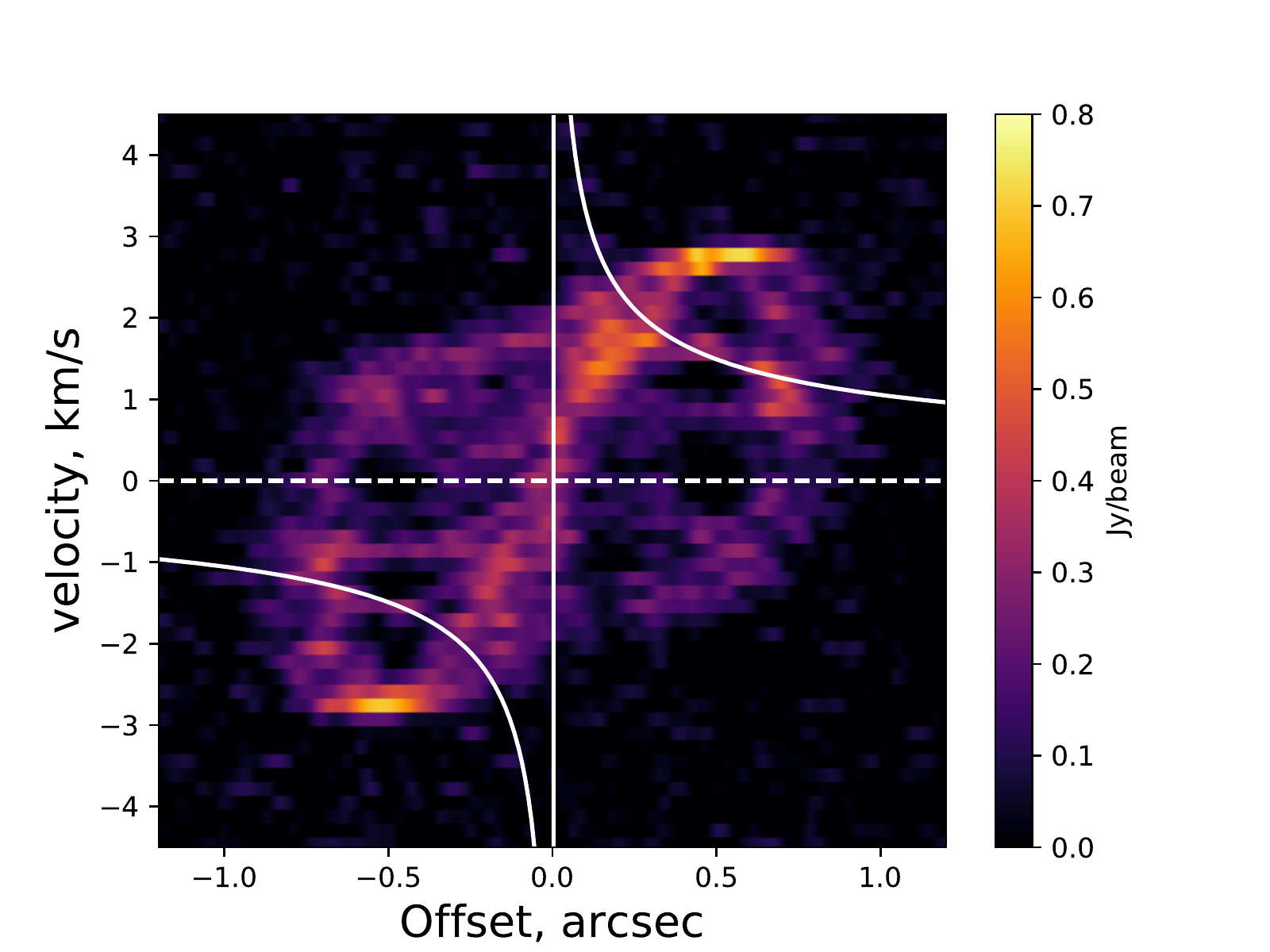}
    \caption{A CI 3P$_2$-3P$_1$ line PV diagram of model B at 45 degree inclination. Comparing with the 3P$_1$-3P$_0$ line image in the central-right panel of Figure \ref{fig:inclination}, the qualitative morphology is similar but the 3P$_2$-3P$_1$ line is a factor 3-4 brighter.}
    \label{fig:CI2to1}
\end{figure}

Comparing the emission as we did in Figure \ref{fig:emission}, the CI 3P$_2$-3P$_1$ is about 4-5 times brighter than the 3P$_1$-3P$_0$ line. This is reflected in the synthetic observations, for example we give an example of a 3P$_2$-3P$_1$ line profile for model B at an inclination of 45 degrees in Figure \ref{fig:CI2to1}. The CI 3P$_1$-3P$_0$ equivalent is given in the centre-right panel of Figure \ref{fig:inclination}. The morphology is qualitatively similar between the two lines, but the 3P$_2$-3P$_1$ is a factor 3-4 brighter. The same applies in the case of the moment 0 map. So our expectation is that the 3P$_2$-3P$_1$ line will not qualitatively differ from the 3P$_1$-3P$_0$ line, but would be easier to detect if awarded time. 

\subsection{Other tracers}
We have prioritised CI since ALMA already has the capability to make resolved observations of this towards candidate evaporating discs in, for example, Orion. Other tracers, particularly CII according to our models, are also expected to provide key insight into externally driven photoevaporative disc winds. However CII is more challenging to observe, with the sensitvity of SOFIA \citep{2012ApJ...749L..17Y} limiting possible targets to nearby low-mass star forming regions with weak UV environments. We will make an assessment of observables (including CII) in weak UV environments in future work.

\section{Summary and Conclusions}
We have provided long needed theoretical predictions of the observational signatures of externally photoevaporating protoplanetary discs in intermediate UV environments ($\sim1000$\,G$_0$)  using CI as a tracer. We have achieved this by producing synthetic observations from photochemical-dynamical models of external disc photoevaporation. CO is a poor tracer of external photoevaporation in intermediate UV environments, since it is dissociated close to the base of the flow and so is only detected where the wind is slow. It is therefore difficult to kinematically detect the wind, though in principle this could be with sufficiently high spectral resolution to detect the small sub-Keplerian rotation \citep{2016MNRAS.457.3593F, 2016MNRAS.463.3616H, 2018A&A...609A..47P, 2019MNRAS.485.3895H}.  The main summary of observing external photoevaporation with CI is as follows \\

1) Unlike CO, the CI 3P$_1$-3P$_0$ line is predominantly optically thin and traces from the sonic surface out to the CII front. The wind is kinematically much easier to detect than CO in CI moment 1 maps, but particularly within CI PV diagrams. Moment 1 maps exhibit large scale twists, which is important as it distinguishes external photoevaporation from other processes that induce twists at smaller radii in the disc. PV diagrams exhibit a range of concentric loops and arcs depending on the inclination, but generally are characterised by velocities higher than that which would result from Keplerian rotation and emission in the two quadrants of PV space where none is found for a purely rotational disc.   \\

2) We develop a simple model of optically thin rings of emission that could be used to infer the parameters of external photoevaporative winds in real systems. This model naturally explains the origin of concentric ellipses in evaporating disc PV diagrams in terms of the radial and vertical components of the flow. \\

3) The CI 3P$_2$-3P$_1$ line is a factor few brighter than the 3P$_1$-3P$_0$ line, but otherwise shows qualitatively similar morphology. At 809\,GHz the 3P$_2$-3P$_1$ line requires ALMA band 10 observations to detect, which are extremely sensitive to the atmospheric conditions.   \\

\section*{Acknowledgements}
{We thank the anonymous reviewer for their comments on the manuscript.}
{We also thank John Ilee, Olja Pani\'{c} and Cathie Clarke for useful discussions.}
TJH is funded by a Royal Society Dorothy Hodgkin Fellowship. JEO is funded by a Royal Society University Research Fellowship.

This work was performed using the DiRAC Data Intensive service at Leicester,
operated by the University of Leicester IT Services, which forms part of the
STFC DiRAC HPC Facility (www.dirac.ac.uk). The equipment was funded by BEIS
capital funding via STFC capital grants ST/K000373/1 and ST/R002363/1 and STFC
DiRAC Operations grant ST/R001014/1. DiRAC is part of the National
e-Infrastructure.

\bibliographystyle{mnras}
\bibliography{molecular}

\appendix

\section{Deriving the parametric ellipse equation}
\label{sec:derive}
We define a circular ring of radius $a$ in a Cartesian co-ordiante system such that the ring occupies the $x-y$ plane, perpendicular to the $z$ axis, with velocity at each point on the ring $(v_R,v_\phi,\pm v_z)$. Now defining a sky co-ordinate system $(X,Y,Z)$, with $\hat{Z}$ the unit vector along the line of sight, so images appear in the $(X,Y)$ plane.

For simplicity, we ignore the position angle of the disc here and only include the effect of inclination. Therefore, the rotation matrix that rotates the disc co-ordinates ($x,y,z$) into the sky co-ordinates $(X,Y,Z)$ is:
\begin{equation}
\begin{pmatrix}
1 & 0 & 0 \\
0 & \cos i & -\sin i \\
0 & \sin i & \cos i
\end{pmatrix}
\end{equation}
where $i$ is the disc's inclination ($i=0$, being face on).  

Now, defining the disc in the disc co-ordinate system to have the following position:
\begin{equation}
\begin{pmatrix}
a\cos E \\
a\sin E \\
0
\end{pmatrix}
\end{equation}
where $E$ is the angle between the radial position and the $x$ axis and ranges between $0$ and $2\pi$. Then the position of the disc in the sky co-ordinates $(X,Y,Z)$ are:
\begin{equation}
X=a\cos(E)
\end{equation}
\begin{equation}
Y=a\cos i \sin(E)
\end{equation}
\begin{equation}
Z=-a\sin i \sin(E)
\end{equation}
Projecting into the $X-Y$ image plane we find the disc's image appears as:
\begin{equation}
X=a\cos(E), \quad\quad Y = a\cos i \sin(E)
\end{equation}
This is the equation for an ellipse with the major axis aligned along $\hat{X}$ of size $a$ and minor axis along $\hat{Y}$ of size $a\cos i$. 

Therefore, we can created PV diagrams by integrating along the $Y$ (minor) axis. Obviously, one is free to rotate the image into the correct North-East co-ordinate system. 

Now to calculate the PV diagram we need to calculate the velocity into the line-of-sight of the observers, i.e. the velocity along the $\hat{Z}$ axis. For a disc rotating counter-clockwise in the $x-y$ plane the cartesian velocity components of the ring are:
\begin{eqnarray}
v_x &=& v_R\cos E-v_\phi\sin E\\
v_y &=& v_R\sin E+v_\phi\cos E\\
v_z &=&\pm v_z
\end{eqnarray}
Therefore, the line-of-sight velocity is given by (using the same rotation matrix as above):
\begin{equation}
v_{\rm los} =  v_R\sin i \sin(E)+v_\phi\sin i \cos (E)\pm v_Z\cos i
\end{equation}

\label{lastpage}

\end{document}